\documentclass[11pt]{article}
\usepackage{amssymb}
\begin{document}
\abovedisplayshortskip 12pt
\belowdisplayshortskip 12pt
\abovedisplayskip 12pt
\belowdisplayskip 12pt

\title{{\bf Supersymmetric Extensions of the Harry Dym Hierarchy}}
\author{J. C. Brunelli$^{a}$, Ashok Das$^{b}$ and Ziemowit Popowicz$^{c}$  \\
\\
$^{a}$ Departamento de F\'\i sica, CFM\\
Universidade Federal de Santa Catarina\\
Campus Universit\'{a}rio, Trindade, C.P. 476\\
CEP 88040-900\\
Florian\'{o}polis, SC\\
Brazil\\
\\
$^{b}$ Department of Physics and Astronomy\\
University of Rochester\\
Rochester, NY 14627-0171\\
USA\\
\\
$^{c}$ Institute of Theoretical Physics\\
University of Wroc\l aw\\
pl. M. Borna 9, 50-205, Wroc\l aw\\
Poland}
\date{}
\maketitle

\begin{center}
{ \bf Abstract}
\end{center}

We study the supersymmetric extensions of the Harry Dym hierarchy of
equations. We
obtain the susy-B extension, the doubly susy-B extension as well as
the $N\!\!=\!\!1$ and the $N\!\!=\!\!2$ supersymmetric extensions
for this system. The $N\!\!=\!\!2$ supersymmetric extension is
particularly interesting, since it leads to new classical integrable
systems in the bosonic limit. We prove the
integrability of these systems through the bi-Hamiltonian formulation
of integrable models
and through the Lax description. We also discuss the supersymmetric extension
of the Hunter-Zheng
equation which belongs to the Harry Dym hierarchy of equations.

\newpage

\section{Introduction:}

Supersymmetric extensions of a number of well know bosonic integrable
models have been studied extensively
in the past. The supersymmetric Korteweg-de Vries (sKdV)
equation~\cite{skdv}, the
supersymmetric nonlinear Schr\"odinger (sNLS) equation~\cite{snls} and
the supersymmetric Two-Boson
(sTB) equation~\cite{stb} represent just a few in this category. A
simple supersymmetric covariantization of
bosonic integrable models, conventionally known as the B supersymmetrization
(susy-B), has also attracted a lot
of interest because of the appearance of such models in string
theories. We  have, for instance, the B extensions of the KdV
(sKdV-B) equation~\cite{skdvb},
the supersymmetric TB (sTB-B) equation~\cite{stbb} and so on. Supersymmetric
extensions of integrable
models using a number $N$ of Grassmann variables greater than
one~\cite{n2skdv} and supersymmetric
construction of dispersionless integrable models~\cite{dispersionless}
have also been studied extensively in the past few years. The extended
supersymmetric models are particularly interesting because, in the
bosonic limit, they yield new classical integrable systems.

A classic bosonic integrable equation, the so called Harry Dym (HD)
equation~\cite{kruskal}, has
attracted much interest recently. The proprieties of
this equation are discussed in detail in
Ref. \cite{hereman}, and we simply emphasize that this equation shares the
properties typical of solitonic
equations, namely, it can be solved by the inverse scattering transform,
it has a bi-Hamiltonian
structure and infinitely many symmetries. In fact, the HD equation is
one of the most exotic
solitonic equations and the hierarchy to which it belongs, has a very
rich structure~\cite{brunelli}.
In this hierarchy we also have nonlocal integrable equations such as
the Hunter-Zheng (HZ)
equation~\cite{hunter}, which arises in the study of massive nematic
liquid  crystals as well as in
the study of shallow water waves. The HD equation, on the other hand,
is  relevant in the study of the Saffman-Taylor
problem which describes the motion of a two-dimensional interface
between a viscous and a non-viscous fluid~\cite{kadanoff}.

An earlier attempt to supersymmetrize the HD equation is discussed in
\cite{liu}. However, this study of $N\!\!=\!\!1$
supersymmetrization introduces a bosonic as well as an independent fermionic
superfield, yielding a pair of coupled equations, and, consequently,
is 
not in the conventional spirit of minimal supersymmetrization. In
this paper we intend to study the question of supersymmetrization of
the HD hierarchy systematically. The paper is organized as follows. In
section 2, we review some of the essential results for the HD
equation and its hierarchy. The simpler susy-B extension (sHD-B) and
the doubly B extension (sHD-BB) of the HD hierarchy as well their
bi-Hamiltonian formulation and
Lax pairs are described in section 3. In section 4, we derive the $N\!\!=\!\!1$
supersymmetric extensions of the HD (sHD) equation. We find that, in
this case, there exist two nontrivial $N\!\!=\!\!1$ extensions. In the
case of one of them, we have a bi-Hamiltonian description (we have not
found a Lax representation yet) while in the second case, we have a
Lax description (we have not found a Hamiltonian structure yet that
satisfies the Jacobi identity). We also describe the supersymmetric
extension for the HZ equation. In section 5, we describe the
$N\!\!=\!\!2$ supersymmetrization of the HD hierarchy which yields
four possibilities and we discuss their properties. We end with a
brief conclusion in section 6.

\section{The Harry Dym Hierarchy:}

The Harry Dym equation
\begin{equation}
w_t=(w^{-1/2})_{xxx}\label{harrydym}\;,
\end{equation}
appears in many disguised forms, namely,
\begin{eqnarray}
v_t &=&{1\over 4}v^3 v_{xxx}\;,\nonumber\\
u_t&=&{1\over4}u^{3/2}u_{xxx}-{3\over 8}u^{1/2}u_xu_{xx}+{3\over
  16}u^{-1/2}u_x^3\label{liuhd}\;,\\
r_t&=&(r_{xx}^{-1/2})_x \;, \nonumber
\end{eqnarray}
where $v=-2^{1/3}w^{-1/2}$, $u=v^2$ and $r_{xx}=w$, respectively. In
this paper,
as in \cite{brunelli}, we will confine ourselves, as much as is
possible, to the form of the HD equation as given in (\ref{harrydym}).

The HD equation is a member of the bi-Hamiltonian
hierarchy of equations given by
\begin{equation}
w_t^{(n+1)}=\mathcal{D}_1 {\delta H_{n+1}\over\delta
w}=\mathcal{D}_2 {\delta H_{n}\over\delta w}\;,  \label{bihamiltonian}
\end{equation}
for $n=-2$, where the bi-Hamiltonian structures are
\begin{eqnarray}
\mathcal{D}_1&=&\partial^3\nonumber\;,\nonumber\\
\mathcal{D}_2&=&w\partial+\partial w\;,
\label{harrydymstructures}
\end{eqnarray}
and the Hamiltonians for the HD equation are
\begin{eqnarray}
H_{-1} &=&
\int  dx\,\left(2w^{1/2}\right)\nonumber\;,\\
H_{-2} &=& \int  dx\left({1\over8}w^{-5/2}w_x^2\right)\label{hdcharges}\;.
\end{eqnarray}
We note here that the second structure in (\ref{harrydymstructures})
corresponds to the centerless Virasoro algebra while
\begin{equation}
{\cal D} = {\cal D}_{2} + c {\cal D}_{1}
\end{equation}
represents the Virasoro algebra with a central charge $c$. We note
also that the recursion operator following from
(\ref{harrydymstructures}),
$R=\mathcal{D}_2\mathcal{D}_1^{-1}$, can be explicitly inverted to yield
\begin{equation}
R^{-1}={1\over 2}\,\partial^3 w^{-1/2} \partial^{-1}w^{-1/2}\;.
\end{equation}
Also, the conserved charges
\begin{eqnarray}
H_{0} &=&- \int  dx\,w \nonumber\;,\\
H_0^{(1)} &=&
\int  dx\,\left(\partial^{-1}w\right)\label{casimir}\;,\\
H_0^{(2)} &=& \int  dx\,\left(\partial^{-2}w\right)\;,\nonumber
\end{eqnarray}
are Casimirs (or distinguished functionals) of the Hamiltonian
operator ${\cal D}_1$ (namely, they are annihilated by the Hamiltonian
structure ${\cal D}_{1}$). As a consequence of this, it is possible to
obtain, in an explicit form, equations from
(\ref{bihamiltonian}) for integers $n$ both positive and negative, i.e.,
$n\in\mathbb{Z}$. As shown in
\cite{brunelli}, for $n>0$, we have three classes of nonlocal
equations. However, in this paper
we will only study the hierarchy associated with the local Casimir
$H_0$ in (\ref{casimir}). In
this way, for $n=1$, we obtain from (\ref{bihamiltonian}), with the
conserved charges
\begin{eqnarray}
H_{1}&=&\int  dx\,{1\over
2}(\partial^{-1}w)^2\nonumber\;,\\\noalign{\vskip 5pt}
H_{2}&=&\int  dx\,{1\over
2}(\partial^{-2}w)(\partial^{-1}w)^2\;,\label{nonlocalh}
\end{eqnarray}
the Hunter-Zheng (HZ) equation
\begin{equation}
w_t=-(\partial^{-2}w)w_x-2(\partial^{-1}w)w\;,\label{hunterzheng}
\end{equation}
which is also an important equation that belongs to the Harry Dym hierarchy.

The integrability of the HD equation (\ref{harrydym}) also follows
from its nonstandard Lax representation
\begin{eqnarray}
L&=&{1\over w}\partial^2\;,\nonumber\\
{\partial L\over\partial
t}&=&-2[B,L]\;,\label{lax}
\end{eqnarray}
where
\begin{equation}
B=\left(L^{3/2}\right)_{\ge2}=w^{-3/2}\partial^3-{3\over
4}w^{-5/2}w_x\partial^2\;.\label{hdlaxb}
\end{equation}
Conserved charges, for $n=1,2,3,\dots$, are obtained from
\begin{equation}
H_{-(n+1)}=\hbox{Tr} L^{2n-1\over 2}\;.\label{negativehs}
\end{equation}
A Lax representation for the HZ equation (\ref{hunterzheng}) is also
known and is given by (\ref{lax}) with
\begin{equation}
B={1\over
  4}(\partial^{-2}w)\partial+{1\over4}\partial^{-1}(\partial^{-2}w)
\partial^2\;.\nonumber\\
\end{equation}
However, in this case, the operator $B$ is not directly related
to $L$, and, consequently, the Lax equation is not of much direct use
(in the construction of conserved charges etc).

\section{The Susy-B Harry Dym (sHD-B, sHD-BB) Equations:}

The most natural generalization of an equation to a supersymmetric one
is achieved simply by working in a superspace. We note, from the HD
equation (\ref{harrydym}), that by a simple dimensional
analysis, we can assign the following canonical dimensions to various
quantities
\begin{equation}
[x]=-1\;,\quad[t]=3\;,\quad\hbox{and}\quad[w]=4\;.\nonumber
\end{equation}
The $N\!\!=\!\!1$ supersymmetric equations are best described
in the superspace parameterized by the coordinates
$z=(x,\theta)$, where  $\theta$ represents the Grassmann
coordinate ($\theta^2=0$). In this space, we can define
\begin{equation}
D={\partial\ \over\partial\theta}+\theta{\partial\ \over\partial
  x}\label{supercovariant}\;,
\end{equation}
representing the supercovariant derivative. From
(\ref{supercovariant}) it follows that
\begin{equation}
D^2=\partial\;,
\end{equation}
which determines the dimension of $\theta$ to be
\begin{equation}
[\theta]=-{1\over2}\;.
\end{equation}
Let us introduce the fermionic superfield
\begin{equation}
W=\psi+\theta w\label{superfield}\;,
\end{equation}
which has the canonical dimension
\begin{equation}
[W]=[\psi]={7\over2}\;.
\end{equation}

A simple supersymmetrization of a bosonic system, conventionally known
as the B supersymmetric (susy-B) extension \cite{skdvb}, is
obtained by simply replacing the bosonic variable $w$, in the original
equation, by 
\begin{equation}
(DW)=w+\theta\psi'\label{dsuperfield}\;,
\end{equation}
where $W$ represents a fermionic superfield. This leads to a manifestly
supersymmetric equation and following this for the case of the
equation (\ref{harrydym}), we obtain the susy-B HD (sHD-B) equation
\begin{equation}
W_t=\partial^2D\Bigl((DW)^{-1/2}\Bigr)\label{shd-b}\;,
\end{equation}
where $W$ is the fermionic superfield (\ref{superfield}).

This system is bi-Hamiltonian with the even Hamiltonian operators
\begin{eqnarray}
\mathcal{D}_1&=&\partial^2\nonumber\;,\nonumber\\
\mathcal{D}_2&=&D(DW)D^{-1}+D^{-1}(DW)D\;,
\end{eqnarray}
and the odd Hamiltonians (which follow from (\ref{hdcharges}) under the
substitution $w\rightarrow (DW)$)
\begin{eqnarray}
H_{-1} &=&\int  dz\,2(DW)^{1/2}\nonumber\;,\\
H_{-2} &=& \int  dz\,{1\over8}(DW)^{-5/2}(DW_x)^2\;.
\end{eqnarray}
The Casimirs of ${\cal D}_1$ can be easily identified with the ones
following from (\ref{casimir}).

The sHD-B equation (\ref{shd-b}) has two possible nonstandard Lax
representations. Let
\begin{equation}
L= (DW)^{-1} D^{4} + c W_{x} (DW)^{-2} D^{3}.
\end{equation}
Then, it can be easily checked that the nonstandard Lax equation
\begin{equation}
\frac{\partial L}{\partial t} = \left[(L^{3/2})_{\geq 3} , L\right],
\end{equation}
leads to the sHD-B equation (\ref{shd-b}) for $c=0,-1$. Here the
projection $()_{\geq 3}$ is defined with respect to the powers of the
supercovariant derivative $D$.

For any given integrable bosonic equation, we can also define a
doubly susy-B extension as follows. Just as we defined  a
superspace in the case of $N=1$ supersymmetry, let us define a
superspace parameterized by $z = (x, \theta_{1},\theta_{2})$,
where $\theta_{1},\theta_{2}$ define two Grasmann coordinates
(anti-comuting and nilpotent, namely,
$\theta_{1}\theta_{2}=-\theta_{2}\theta_{1}$, $\theta_{1}^{2} =
\theta_{2}^{2}=0$). In this case, we can define two supercovariant
derivatives
\begin{eqnarray}
D_{1} & = & \frac{\partial}{\partial \theta_{1}} + \theta_{1}
  \frac{\partial}{\partial x}\;,\nonumber\\
D_{2} & = & \frac{\partial}{\partial \theta_{2}} + \theta_{2}
  \frac{\partial}{\partial x}\;,\label{n2der}
\end{eqnarray}
which satisfy
\begin{equation}
D_{1}^{2} = D_{2}^{2} = \partial\;,\quad D_{1}D_{2} + D_{2}D_{1} = 0.
\end{equation}
Such a superspace naturally defines a system with $N=2$
supersymmetry. Let us consider a bosonic superfield, $W$, in this
space which will have the expansion (we denote it by the same symbol
as in the case of $N=1$)
\begin{equation}
W=w_0+\theta_1\chi+\theta_2\psi+\theta_2\theta_1w_1\label{n2superfield}\;.
\end{equation}
Then, we can simply replace the bosonic variable in the original
equation by $(D_{1}D_{2}W)$ which leads to the doubly susy-B extension
of a given equation. For the HD equation (\ref{harrydym}), this leads
to
\begin{equation}
W_{t} = \partial D_{1}D_{2}
\left((D_{1}D_{2}W)^{-1/2}\right)\;,\label{shd-bb}
\end{equation}
which defines the sHD-BB equation. This procedure can, of course, be
generalized to any $N$ extended supersymmetry and we do not pursue
this any further. We simply point out that eq. (\ref{shd-bb}) is
bi-Hamiltonian, as we would expect. For example, it is Hamiltonian
with
\begin{equation}
H = \int dz\, (D_{1}D_{2}W_{x})^{2} (D_{1}D_{2}W)^{-5/2}\;,
\end{equation}
and
\begin{eqnarray}
{\cal D} & = & - \partial W \partial^{-2} D_{1}D_{2} - D_{1}D_{2}
\partial^{-2} W \partial + D_{1} \partial^{-1} W D_{2} - D_{2} W
\partial^{-1} D_{1}\nonumber\\
 &  & + D_{1} D_{2} \partial^{-1} W - W D_{1}D_{2}
\partial^{-1} + D_{1} W \partial^{-1} D_{2} - D_{2} \partial^{-1} W
D_{1}\;.
\end{eqnarray}
The second Hamiltonian structure can also be easily obtained.

\section{The Supersymmetric $N\!\!=\!\!1$ Harry Dym (sHD) and
  Hunter-Zheng (sHZ) Equations:}

As we have seen, the susy-B extension of a system is a very simple
supersymmetrization. However, to obtain nontrivial
supersymmetrizations, we can follow one of the following two 
approaches. In this section, we will discuss $N\!\!=\!\!1$
supersymmetrization of the system and correspondingly, it is
appropriate to work in the superspace defined in
(\ref{supercovariant})--(\ref{superfield}).

With the superfield (\ref{superfield}) as our basic variable, the
first approach is to write the most general
local equation in superspace which is consistent with  all canonical
dimensions and which reduces
to (\ref{harrydym}) in the bosonic limit. This involves a
free parameter and the equation takes the form
\begin{eqnarray}
W_{t} & = & \frac{1}{8}\left[-8 (5a -2) W_{xxx} (DW)^{-3/2} + 2 (65a -
 6) W_{x} (DW_{xx}) (DW)^{-5/2}\right.\nonumber\\
 &  & \quad + 30 (5a + 2) W_{xx} (DW_{x}) (DW)^{-5/2} - 15 (21a + 2)
 W_{x} (DW_{x})^{2} (DW)^{-7/2}\nonumber\\
 &  & \quad + W\left\{8(5a - 6) (DW_{xxx}) (DW)^{-5/2} + 10 (a - 6)
 W_{xxx}W_{x} (DW)^{-7/2}\right.\nonumber\\
 &  & \quad + 35 (6 - a) W_{xx}W_{x} (DW_{x}) (DW)^{-9/2} + 40 (6 -
 7a) (DW_{xx}) (DW_{x}) (DW)^{-7/2}\nonumber\\
 & & \quad \left.\left. + 105 (3a - 2) (DW_{x})^{3}
 (DW)^{-9/2}\right\}\right]\;,\label{1}
\end{eqnarray}
where $a$ is the arbitrary parameter. In the case of the HD equation,
it is  possible
to supersymmetrize the two Hamiltonian structures in
(\ref{bihamiltonian}), which is easily seen from the fact that the
second Hamiltonian structure is the centerless Virasoro algebra. Thus,
the supersymmetrized Hamiltonian structures follow to be
\begin{eqnarray}
\mathcal{D}_1&=&D\partial^2\nonumber\;,\nonumber\\
\mathcal{D}_2&=& \frac{1}{2}\left[W\partial+2\partial
  W+(DW)D\right]\;.\label{susystructures}
\end{eqnarray}
Requiring eq. (\ref{1}) to be bi-Hamiltonian with respect to
(\ref{susystructures}), namely, requiring
\begin{equation}
W_t=\mathcal{D}_1 {\delta H_{-1}\over\delta
W}=\mathcal{D}_2 {\delta H_{-2}\over\delta W}\;,\label{susybihamiltonian}
\end{equation}
determines the parameter to be $a = 6$.
The Hamiltonians in (\ref{susybihamiltonian}), in this case have the
forms
($dz=dx\,d\theta$ with $\int d\theta=0$ and $\int d\theta\,\theta=1$)
\begin{eqnarray}
H_{-1} &=&
\int  dz\,2W(DW)^{-1/2}\nonumber\;,\\
H_{-2} &=& \int
dz\,{1\over8}\left[W_x(DW_x)(DW)^{-5/2}-15WW_xW_{xx}(DW)^{-7/2}\right]\;,
\end{eqnarray}
and the $N\!\!=\!\!1$ sHD equation assumes the simple form
\begin{equation}
W_t=D\partial^2\left(2(DW)^{-1/2}-3WW_x(DW)^{-5/2}\right)\;.\label{shd}
\end{equation}
It is worth noting here that this equation differs from the sHD-B
equation (\ref{shd-b}) in the presence of the second term inside the
parenthesis on the right hand side, which vanishes in the bosonic
limit. (We would like to point out parenthetically that we do not
generate the sHD-B equation in this approach because of our
requirement that the equation be bi-Hamiltonian with respect to the
structures in (\ref{susystructures}).)

It is easy to check that the Hamiltonian $H_{-1}$ is a Casimir of
${\cal D}_2$  and the conserved charge
\begin{equation}
H_0=-\int dz\,W
\end{equation}
is a Casimir of ${\cal D}_1$. Furthermore, the Hamiltonian structure ${\cal
  D}_2$ can be written in the form
\begin{equation}
{\cal D}_2= \frac{1}{2} (DW)^{1/2}D(1+X)(DW)^{1/2}\;,
\end{equation}
where
\begin{equation}
X\equiv {3\over2}\left(D{W\over(DW)}D^{-1}+D^{-1}{W\over(DW)}D\right)D\;,
\end{equation}
and therefore can be formally inverted. Thus, in this case also the
associated recursion operator has a formal inverse.

It can be easily checked that the following charges
\begin{eqnarray}
H_{1} &=&
\int  dz\,{1\over4}(D^{-1}W)(D^{-2}W)\nonumber\;,\\
H_{2} &=& \int  dz\,{1\over2}(D^{-1}W)(D^{-2}W)(D^{-3}W)\;,
\end{eqnarray}
are conserved and reduce to (\ref{nonlocalh}) in the bosonic limit. From
\begin{equation}
W_t=\mathcal{D}_1 {\delta H_{2}\over\delta
W}=\mathcal{D}_2 {\delta H_{1}\over\delta W}\;,  \label{hzbihamiltonian}
\end{equation}
we obtain the $N\!\!=\!\!1$ supersymmetric HZ (sHZ) equation
\begin{equation}
W_t=-{3\over2}W(D^{-1}W)-W_x(D^{-3}W)-{1\over2}(DW)(D^{-2}W)\;.
\end{equation}
Both the sHD and the sHZ equations are bi-Hamiltonian systems and the
infinite set of commuting conserved charges can be constructed
recursively. As a result, they decribe supersymmetric integrable
systems.

The second approach to finding a nontrivial $N\!\!=\!\!1$
supersymmetrization of the HD equation is to start with the Lax
operator in (\ref{lax}) and generalize it to superspace. Let us start
with the most general Lax operator involving non-negative powers of $D$,
\begin{equation}
L=a_{0}^2D^4+\alpha_1D^3+a_1D^2+\alpha_2D+a_2\;,\label{ansatz}
\end{equation}
with the identification
\begin{equation}
a_0=(DW)^{-1/2}\;,
\end{equation}
where Roman coefficients are bosonic and Greek ones are
fermionic. It is easy to verify that, in this case, there are only
three projections, $()_{\geq 0,1,3}$ (with respect to powers of $D$),
that can lead to a consistent Lax
equation. Using this ansatz,  we have not yet been able to
obtain the sHD equation (\ref{shd-b}) using fractional powers of the
Lax  operator (\ref{ansatz}). The Lax pair for this system, therefore, remains
an open question.

On the other hand, when
\begin{equation}
\alpha_{1} = c W_{x} (DW)^{-2},\quad a_{1} = a_{2} = 0 = \alpha_{2},
\end{equation}
where $c$ is an arbitrary parameter, the nonstandard Lax equation
\begin{equation}
{\partial L\over\partial t}=[(L^{3/2})_{\ge3},L]\;,
\end{equation}
yields consistent equations only for $c=0,-1,-\frac{1}{2}$. As we have
pointed out in the last section, for the values of the parameter,
$c=0,-1$, we have the sHD-B equation. The third choice of the
parameter, therefore, leads to a new nontrivial $N\!\!=\!\!1$
supersymmetrization of the HD equation. Namely, with
\begin{equation}
L = (DW)^{-1} D^{4} - \frac{1}{2} W_{x} (DW)^{-2} D^{3}\;,\label{second}
\end{equation}
the Lax equation
\begin{equation}
\frac{\partial L}{\partial t} = \left[(L^{3/2})_{\geq 3}, L\right]\;,
\end{equation}
leads to a second $N=1$ supersymmetrization of the HD equation of the
form
\begin{eqnarray}
W_{t} & = & \frac{1}{16}\left[8D^{5} ((DW)^{-1/2}) - 3D(W_{xx}W_{x}
  (DW)^{-5/2})\right.\nonumber\\
 &  & \quad\left. + \frac{3}{4} (DW_{x})^{2} W_{x} (DW)^{-7/2} -
  \frac{3}{4} D^{-1}\left((DW_{x})^{3}
  (DW)^{-7/2}\right)\right]\;.\label{secondequation}
\end{eqnarray}
This is manifestly a nonlocal susy generalization in the variable $W$
which, however, is a completely local equation in the variable $(DW)$.

Since this system of equations has a Lax description, it is integrable
and the conserved
charges can be calculated in a standard manner and the first few
charges take the forms
\begin{eqnarray}
H_{1} & = & \int dz\, W_{x} (DW_{x}) (DW)^{-5/2}\;,\\
H_{2} & = & \int dz\, W_{x}\left[16 (DW_{xxx})(DW)^{-7/2} - 84
  (DW_{xx})(DW_{x})(DW)^{-9/2} + 77
  (DW_{x})^{3}(DW)^{-11/2}\right]\;,\nonumber
\end{eqnarray}
and so on. However, we have not yet succeeded in finding a
Hamiltonian structure
which satisfies Jacobi identity (it is clear that the Hamiltonian
structure is nonlocal, since the Hamiltonian is local).

\section{The $N\!\!=\!\!2$ Supersymmetric Harry Dym Hierarchy:}

The most natural way to discuss the $N=2$ supersymmetric
extension of the HD equation is in the $N=2$ superspace
introduced earlier in (\ref{n2der})--(\ref{n2superfield}).
Looking at the bosonic superfield $W$ in (\ref{n2superfield}), we
note that it has two bosonic components as well as two fermionic
components. In the bosonic limit, when we set the fermions to
zero, the $N=2$ equation would reduce to two bosonic equations.
Since we have only the single HD equation (\ref{harrydym}) to
start with, the construction of such a system is best carried out
in the Lax formalism. This also brings out the interest in such
extended supersymmetric systems, namely, they lead to new bosonic
integrable systems in the bosonic limit.

As in (\ref{ansatz}), let us consider the most general $N=2$ Lax
operator which contains differential operators in this
superspace of the following form (taking a more general Lax involving
only differential operators does not lead to equations which reduce to
the HD equation),
\begin{eqnarray}
L & = & W^{-1}\partial^{2} + (D_{1}W^{-1}) (\kappa_{1} D_{1} + \kappa_{2}
D_{2}) \partial + (D_{2}W^{-1}) (\kappa_{3} D_{1} + \kappa_{4} D_{2})
\partial\nonumber\\
 &  & \quad + \left(\kappa_{5} (D_{1}D_{2}W) W^{-2} + \kappa_{6}
   (D_{1}W)(D_{2}W) W^{-3}\right) D_{1} D_{2}\;,\label{n2lax}
\end{eqnarray}
where $\kappa_{i}, i=1,2,\cdots , 6$ are arbitrary constant
parameters. The $N=2$ supersymmetry corresponds to an internal $O(2)$
invariance that rotates $\theta_{1}\rightarrow \theta_{2},
\theta_{2}\rightarrow -\theta_{1}$ and correspondingly
$D_{1}\rightarrow D_{2}, D_{2}\rightarrow - D_{1}$ (thereby rotating
the  fermion components of the superfield
into each other). This invariance, imposed on the Lax operator,
identifies
\begin{equation}
\kappa_{4} = \kappa_{1},\quad \kappa_{3} = -\kappa_{2}\;.
\end{equation}

Using the computer algebra program REDUCE \cite{hearn} and the
special package SUSY2 \cite{susy2}, we are able to study
systematically the hierarchy of equations following from the Lax
equation
\begin{equation}
\frac{\partial L}{\partial t} = \left[(L^{3/2})_{\geq 2},
  L\right]\;.\label{n2laxequation}
\end{equation}
Here, the projection (which is the highest consistent projection as is
also the case with the bosonic HD equation in (\ref{hdlaxb})) is
understood as  follows. Let us recall that a
general pseudodifferential operator in $N=2$ superspace has the form
\begin{equation}
P = \sum_{n=-\infty}^{n=\infty} \left(P_{0}^{n} + P_{1}^{n} D_{1} +
  P_{2}^{n} D_{2} +  P_{3}^{n} D_{1}D_{2}\right) \partial^{n}\;.
\end{equation}
For such a pseudodifferential operator, the projection in
(\ref{n2laxequation}) is defined as
\begin{eqnarray}
P_{\geq 2} & = & P_{3}^{0} D_{1}D_{2} + \left(P_{1}^{1} D_{1} +
  P_{2}^{1} D_{2} + P_{3}^{1} D_{1}D_{2}\right)\partial\nonumber\\
 &  & \quad + \sum_{n\geq 2} \left(P_{0}^{n} + P_{1}^{n} D_{1} +
  P_{2}^{n} D_{2} + P_{3}^{n} D_{1}D_{2}\right) \partial^{n}\;.
\end{eqnarray}

The consistency of the equation (\ref{n2laxequation}) leads to four
possible solutions for the values of the arbitrary parameters
\begin{enumerate}
\item $\kappa_{1} = \kappa_{2} = \kappa_{5} = \kappa_{6} = 0\;,$
\item $\kappa_{2} = 0$, $\kappa_{1} = \kappa_{5} = -
  \frac{\kappa_{6}}{2} = 1\;,$
\item $\kappa_{2} = \kappa_{5} = \kappa_{6} = 0$, $\kappa_{1} =
  \frac{1}{2}\;,$
\item $\kappa_{2} = 0$, $\kappa_{1} = \kappa_{5} = \frac{1}{2},
  \kappa_{6} = \frac{3}{4}\;.$
\end{enumerate}
We will now discuss the various cases separately in some detail.

The
first and the second cases can be discussed together since they lead
to the same dynamical equation. Namely, in this case, the two Lax
operators take the forms
\begin{eqnarray}
L^{(1)} & = & W^{-1} \partial^{2}\;,\nonumber\\
L^{(2)} & = & W^{-1} \partial^{2} + (D_{1}W^{-1}) D_{1}\partial +
(D_{2}W^{-1}) D_{2}\partial - (D_{1}D_{2}W^{-1}) D_{1}D_{2}\nonumber\\
 & = & - D_{1}D_{2} W^{-1} D_{1}D_{2}\;.
\end{eqnarray}
It can be checked that both these Lax operators lead to the same
dynamical equation which is nothing other than the sHD-BB equations we
have discussed earlier and, therefore, we do not study this any
further.

For the third choice of parameters, the Lax operator can be written in
the simple form
\begin{eqnarray}
L^{(3)} & = & W^{-1} \partial^{2} +
\frac{1}{2}\left((D_{1}W^{-1})D_{1} + (D_{2}W^{-1})D_{2}\right)
\partial\nonumber\\
 & = & \frac{1}{2} \left(D_{1} W^{-1} D_{1} + D_{2} W^{-1}
D_{2}\right) \partial\;.\label{3lax}
\end{eqnarray}
The Lax equation (\ref{n2laxequation}), in this case, leads to a
nontrivial $N=2$ supersymmetric HD equation of the form
\begin{eqnarray}
W_{t} & = & \frac{1}{64}\left[2 (W^{-1/2})_{xxx} -12 (D_{1}W_{xx})
  (D_{1}W) W^{-5/2} - 12 (D_{2}W_{xx}) (D_{2}W)
  W^{-5/2}\right.\nonumber\\
 &  & \quad + 36 (D_{1}W_{x}) (D_{1}W) W_{x} W^{-7/2} + 36
  (D_{2}W_{x})(D_{2}W) W_{x} W^{-7/2}\nonumber\\
 &  & \quad \left. + 6 (D_{1}W)(D_{2}W)(D_{1}D_{2}W_{x}) W^{-7/2}
- 9 (D_{1}W)(D_{2}W) (D_{1}D_{2}W) W_{x}
  W^{-9/2}\right]\;.\label{3equation}
\end{eqnarray}
In the bosonic sector, where we set all the fermions to zero so that
(see (\ref{n2superfield}))
\begin{equation}
W = w_{0} + \theta_{2}\theta_{1} w_{1}\;,
\end{equation}
the equation (\ref{3equation}) reduces to
\begin{eqnarray}
w_{0,t} & = & \frac{1}{2} (w_{0}^{-1/2})_{xxx}\;,\nonumber\\
w_{1,t} & = & \frac{1}{64}\left[-16 w_{1,xxx}w_{0}^{-3/2} + 96
  w_{1,xx}w_{0,x}w_{0}^{-5/2} + 72
  w_{1,x}w_{0,xx}w_{0}^{-5/2}\right.\nonumber\\
 &  & \quad -258 w_{1,x}w_{0,x}^{2}w_{0}^{-7/2} - 6
  w_{1,x}w_{1}^{2}w_{0}^{-7/2} + 9
  w_{1}^{3}w_{0,x}w_{0}^{-9/2}\nonumber\\
 &  & \quad \left.-108 w_{1}w_{0,xx}w_{0,x}w_{0}^{-9/2} + 219
  w_{1}w_{0,x}^{3}w_{0}^{-9/2}\right]\;.\label{3components}
\end{eqnarray}
The first of the equations in (\ref{3components}) is, of course, the
HD equation (\ref{harrydym}), but is decoupled from the second
component. Consequently, even though this set of equations
represents a new integrable system, it is not very interesting. Let us
note that we can reduce the $N=2$ supersymmetry of this system to
$N=1$ supersymmetry in the following way. Let us define
\begin{equation}
W (x,\theta_{1},\theta_{2}) = U(x,\theta_{1}) + \theta_{2} F
(x,\theta_{1})\;,\label{reduction}
\end{equation}
and set the fermionic superfield $F(x,\theta_{1})=0$. This would,
therefore, make the superfield $W$ independent of the Grassmann
coordinate $\theta_{2}$ leaving us with $N=1$ supersymmetry. Under
such a reduction, it is straightforward to see that the Lax operator
(\ref{3lax}) and the equation (\ref{3equation}) go over to the ones in
(\ref{second}) and (upto multiplicative factors) the corresponding
equation (\ref{secondequation}) with the identification
\begin{equation}
\theta_{1} = \theta,\quad U (x,\theta_{1}) = (DW(x,\theta))\;.
\end{equation}
The conserved charges for this system can be obtained from the Lax
operator $L^{(3)}$ in a standard manner, but we do not go into the
details of this.

The fourth case is probably the most interesting of all. Here, the Lax
operator takes the form
\begin{eqnarray}
L^{(4)} & = & W^{-1} \partial^{2} + \frac{1}{2}
\left((D_{1}W^{-1})D_{1} +
(D_{2}W^{-1})D_{2}\right)\partial\nonumber\\
 &  & \quad - W^{-1/2} (D_{1}D_{2}W^{-1/2}) D_{1}D_{2}\nonumber\\
 & = & - \left(W^{-1/2}D_{1}D_{2}\right)^{2}\;.\label{4lax}
\end{eqnarray}
Interestingly enough, this Lax operator possesses two nontrivial
square roots, namely,
\begin{eqnarray}
L_{1}^{1/2} & = & iW^{-1/2} D_{1}D_{2}\;,\nonumber\\
L_{2}^{1/2} & = & W^{-1/2} \partial +
\frac{1}{2}\left[(D_{1}W^{-1/2})D_{1} + (D_{2}W^{-1/2})D_{2} -
  (W^{-1/2})_{x}\right] + \cdots\;.\label{squareroot}
\end{eqnarray}
We note here that a similar situation also arises in the study of the
$N=2$ sKdV hierarchy \cite{n2susy} (for the case of the parameter
$a=4$). In  such a
case, the general hierarchy of equations can be obtained from the Lax
equation
\begin{equation}
\frac{\partial L}{\partial t_{n}} = \left[\left(L_{1}^{n/2}
  L_{2}^{1/2}\right)_{\geq 2}, L\right]\;,\label{4laxequation}
\end{equation}
where $n=0,1,2,\cdots$. For example, the first two flows of the
hierarchy take the forms
\begin{eqnarray}
W_{t_{1}} & = & \frac{1}{8}\left[-4(D_{1}D_{2}W_{x}) W^{-1} + 6
  (D_{1}D_{2}W)W_{x} W^{2}\right.\nonumber\\
 &  & \quad\left.+ 6 \left((D_{1}W) (D_{2}W_{x}) - 6 (D_{2}W)
  (D_{1}W_{x})\right) W^{-2} - 15 (D_{1}W)(D_{2}W)W_{x}
  W^{-3}\right]\;,\nonumber\\
W_{t_{2}} & = & \frac{1}{16}\left[8(W^{-1/2})_{xxx} - 6
  (D_{1}D_{2}W)(D_{1}D_{2}W_{x}) W^{-5/2} + 9 (D_{1}D_{2}W)^{2}
  W_{x}W^{-7/2}\right.\nonumber\\
 &  & \quad + 3 \left((D_{1}W)(D_{1}W_{xx}) +
  (D_{2}W)(D_{2}W_{xx})\right) W^{-5/2}\nonumber\\
 &  & \quad - 9\left((D_{1}W)(D_{1}W_{x}) +
  (D_{2}W)(D_{2}W_{x})\right)W_{x} W^{-7/2}\nonumber\\
 &  & \quad \left. + 9 ((D_{1}W)(D_{2}W)(D_{1}D_{2}W))_{x} W^{-7/2} -
  36  (D_{1}W)(D_{2}W)(D_{1}D_{2}W) W_{x}
  W^{-9/2}\right]\,.\nonumber\\
  \noalign{\vspace{-7pt}}\label{4equation}
\end{eqnarray}
 In the bosonic sector, the second equation in
(\ref{4equation}) gives
\begin{eqnarray}
w_{0,t_{2}} & = & \frac{1}{16}\left[8 (w_{0}^{-1/2})_{xxx} - 6
  w_{1,x}w_{1}w_{0}^{-5/2} + 9
  w_{1}^{2}w_{0,x}w_{0}^{-7/2}\right]\;,\nonumber\\
w_{1,t_{2}} & = & \frac{1}{32}\left[-8w_{1,xxx}w_{0}^{-3/2} +
  48\left(w_{1,x}w_{0,x}\right)_{x} w_{0}^{-5/2} - 144
  w_{1,x}w_{0,x}^{2}w_{0}^{-7/2}\right.\nonumber\\
 &  & \quad - 6 w_{1,x}w_{1}^{2}w_{0}^{-7/2} + 9
  w_{1}^{3}w_{0,x}w_{0}^{-9/2} + 12
  w_{1}w_{0,xxx}w_{0}^{-5/2}\nonumber\\
 &  & \quad \left. -126 w_{1}w_{0,xx}w_{0,x}w_{0}^{-7/2} + 177
  w_{0,x}^{3}w_{1}w_{0}^{-9/2}\right]\;.
\end{eqnarray}
This is a new bosonic system of coupled equations, which reduces on
setting $w_{1}=0$ to the HD equation and is integrable.

The conserved charges for this last case of $N=2$ supersymmetrization
can be constructed as follows.
\begin{eqnarray}
H_{1} & = & \int dz\,{\rm sRes} L_{2}^{1/2} = \int
dz\,(D_{1}W)(D_{2}W)W^{-5/2}\;,\nonumber\\
H_{2} & = & \int dz\, {\rm sRes} (L_{1}^{1/2}L_{2}^{1/2})\nonumber\\
 & = & \int dz\left[3(D_{1}D_{2}W)^{2} W^{-3} + 3 W_{x}^{2} W^{-2}
  \right.\nonumber\\
 &  & \quad\left. + 2\left((D_{1}W)(D_{1}W_{x}) +
  (D_{2}W)(D_{2}W_{x})\right)W^{-3} + (D_{1}W)(D_{2}W)(D_{1}D_{2}W)
  W^{-4}\right]\;,\nonumber\\
H_{3} & = & \int dz\, {\rm sRes} L_{2}^{3/2}\nonumber\\
 & = & \int dz\left[128 (D_{1}D_{2}W_{x})W_{x}W^{-7/2} - 40
(D_{1}D_{2}W)^{3} W^{-9/2} + \cdots \right]\;,
\end{eqnarray}
where $dz = dxd\theta_{1}d\theta_{2}$ and ``${\rm sRes}$'' is
defined as the coefficient of the $D_{1}D_{2}\partial^{-1}$ term
in the pseudodifferential operator. We can also perform the $N=1$
reduction of this system. Requiring that the superfield $W$ has no
dependence on $\theta_{2}$, it is clear from the form of the Lax
operator in (\ref{4lax}) that it reduces to the one involving the
second $N=1$ supersymmetrization (just as $L^{(3)}$ does).

\section{Conclusions:}

In this paper, we have studied the question of
supersymmetrization of the Harry Dym hierarchy systematically. We
have used the simpler B supersymmetrization to derive the sHD-B
and sHD-BB systems. The analysis of the nontrivial $N=1$
supersymmetrization leads to two such integrable systems. One has
a natural bi-Hamiltonian description for which we have not been
able to find the Lax description. On the other hand, the second
has a natural Lax description for which we have not yet found a
Hamiltonian structure that satisfies the Jacobi identity. Both
these systems are integrable. The $N=2$ supersymmetrization from
the Lax approach yields four possible Lax operators. Two of these
describe the sHD-BB system while the other two give nontrivial
$N=2$ supersymmetric extensions. In the bosonic limit, one of
them leads to the HD equation decoupled from the second component
while the other genuinely gives a coupled two component system of
equations that is integrable.

\section*{Acknowledgments}

One of us (ZP) would like to thank the high energy theory group at the
University of Rochester for hospitality where this work was done. This
work  was supported in part by US DOE grant
no. DE-FG-02-91ER40685 as well as by NSF-INT-0089589.

\end{document}